# Implementation of PFC and RCM for RoCEv2 Simulation in OMNeT++

Qian Liu*, Robert D. Russell*, Fabrice Mizero†, Malathi Veeraraghavan†, John Dennis‡, Benjamin Jamroz‡

*Department of Computer Science, University of New Hampshire, Durham, NH, USA
†Dept. of Electrical & Computer Engineering, University of Virginia, Charlottesville, VA, USA
‡National Center for Atmospheric Research, Boulder, CO, USA
Email: *{qga2, rdr}@unh.edu    †{fm9ab, mv5g}@virginia.edu    ‡{dennis, jamroz}@ucar.edu

*Abstract*—As traffic patterns and network topologies become more and more complicated in current enterprise data centers and TOP500 supercomputers, the probability of network congestion increases. If no countermeasures are taken, network congestion causes long communication delays and degrades network performance. A congestion control mechanism is often provided to reduce the consequences of congestion. However, it is usually difficult to configure and activate a congestion control mechanism in production clusters and supercomputers due to concerns that it may negatively impact jobs if the mechanism is not appropriately configured. Therefore, simulations for these situations are necessary to identify congestion points and sources, and more importantly, to determine optimal settings that can be utilized to reduce congestion in those complicated networks. In this paper, we use OMNeT++ to implement the IEEE 802.1Qbb Priority-based Flow Control (PFC) and RoCEv2 Congestion Management (RCM) in order to simulate clusters with RoCEv2 interconnects.

*Keywords*-Congestion Control; Flow Control; OMNeT++; RoCEv2

## I. INTRODUCTION

Most networks have congestion points or bottlenecks. Network congestion usually occurs when the total demand for a link is greater than the capacity of the link. In lossless networks such as InfiniBand [1] and RDMA over Converged Ethernet version 2 (RoCEv2) [2], the impact of congestion can be severe and can cause long communication delay.

In current enterprise data centers and TOP500 [3] supercomputers that employ InfiniBand, network congestion is inevitable as traffic patterns and network topology become complicated. Jobs running on supercomputers have experienced significant variation in latency due to congestion [4, 5]. However, huge clusters seldom utilize congestion control due to concerns that, if they are not appropriately configured, such mechanisms may negatively impact production jobs. Therefore, simulations for those congestion situations, especially situations in today's huge clusters that form supercomputers, are needed to identify congestion points and sources without modifying or rewriting the applications experiencing congestion, and more importantly, to determine optimal settings for the congestion control mechanism.

OMNeT++ [6] is an extensible, modular, component-based C++ network event simulator. Network components and basic elements can be organized in modules, which can be connected via communication gates. An open-source OMNeT++ simulation model [7] released by Mellanox implements the InfiniBand credit-based flow control mechanism and Quality-of-Service (QoS) that supports arbitration among different Virtual Lanes (VLs). Integration of InfiniBand congestion control into this model was discussed in [8], but that implementation is not yet open-source.

In this paper, we extend the current InfiniBand OMNeT++ model [7] released by Mellanox (abbr. the MLNX model) with Priority-based Flow Control (PFC, IEEE 802.1Qbb [9]) and RoCEv2 Congestion Management (RCM) [2] in order to simulate RoCEv2 clusters. The reason we based our work on the MLNX model is that latest Mellanox CAs with 2 ports can be configured to run both ports as InfiniBand, both as RoCE, or to run one port in each mode. Therefore, internally they must have a lot of shared hardware/firmware and be based on nearly identical architectures. Additionally, the MLNX simulation model avoids InfiniBand details by dealing abstractly with data arbitration, transmission and forwarding based on QoS, concepts that are very close to, and can be reused in, RoCEv2.

## II. BACKGROUND

The InfiniBand architecture [1] defines the 4 lowest layers of the OSI reference stack. RoCE [10] preserves InfiniBand's verbs interface and its transport and network layers, but utilizes Ethernet's link and physical layers as well as their management infrastructure. Its packets are not routable. RoCEv2 [2] preserves InfiniBand's verbs interface and transport layer, but utilizes standard IP layer and Ethernet's link and physical layers. RoCEv2 packets can be routed.

TABLE I: Flow/Congestion Control Mechanisms in InfiniBand, RoCE and RoCEv2. (O means "can be utilized", × means "cannot be utilized", √ means "defined in the spec")

|        | InfiniBand | RoCE | RoCEv2 |
|--------|:----------:|:----:|:------:|
| IB CFC | √          | ×    | ×      |
| PAUSE  | ×          | O    | O      |
| PFC    | ×          | O    | O      |
| IB CC  | √          | ×    | ×      |
| QCN    | ×          | O    | ×      |
| RCM    | ×          | ×    | √      |

Table I compares the underlying flow control and congestion control mechanisms implemented in InfiniBand, RoCE and RoCEv2. Lossless behavior in InfiniBand is achieved by its Credit-based Flow Control (CFC) mechanism. While RoCE [10] doesn't specifically require losslessness, its performance suffers if link layer flow control mechanism is not provided [11]. The RoCEv2 specification [2] doesn't define a mechanism to achieve losslessness but it requires such behavior from the network, link, and physical layers below its InfiniBand transport layer. Losslessness in RoCEv2 can be achieved through the use of a link-layer flow control mechanism such as the Priority-based Flow Control (PFC) defined in IEEE 802.1Qbb [9], which extends the IEEE 802.3x PAUSE semantics to apply to multiple classes of service (Virtual Links, VLs).

With the introduction of link-layer flow control, packets are not longer dropped due to buffer overflow. Instead, a packet is simply not sent on the link unless the other end of the link has buffer space to receive it. This may cause the sending side buffer to fill, so that congestion spreads upstream. The InfiniBand congestion control mechanism is defined in [1] as a way to reduce congestion and its spreading. The RoCE specification [10] doesn't define any congestion countermeasures. However, RoCEv2 [2] provides its own level 3 end-to-end "RoCEv2 Congestion Management" (RCM) mechanism [2] that seeks to alleviate congestion. Network Elements (NEs, routers or switches) and Channel Adapters (CAs, host interfaces) play different roles in this mechanism. NEs are responsible for detecting congestion and notifying destination end points. For congestion notification, RCM relies on the IP mechanism defined in Explicit Congestion Notification (ECN) [12], in which NEs mark exiting packets involved in congestion using the ECN field in the IP header. A CA that receives a packet with a value of '11' in its *IP.ECN* field is responsible for notifying the packet source about congestion by returning a Congestion Notification Packet (CNP). A CA that receives a CNP is responsible for reducing its packet injection rate.

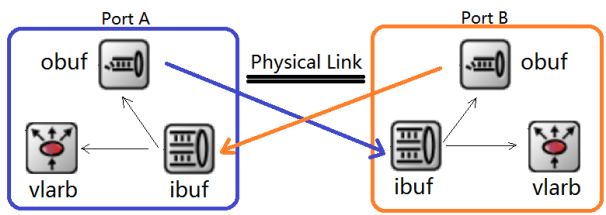

Figure. 1: Credit-based Flow Control (CFC) in the MLNX model

## III. RoCE Priority-based Flow Control (PFC) Implementation

Figure 1 illustrates the credit-based flow control abstractions implemented in the current MLNX simulation model. In InfiniBand, one credit signifies 64 bytes of available receive buffer space. Whenever there is a change in credits on Port A, for instance, its *obuf* module sends a flow control packet across the physical link to the *ibuf* module in port B, which notifies its *vlarb* module about the number of available credits in port A that can be used to receive data packets from port B, and also notifies its *obuf* module about its local available credits, allowing the *obuf* module in port B to send a flow control packet to port A when necessary. Flow control packets should be sent frequently to prevent any possible long latency in data transmission. This frequency is configurable in the MLNX model.

Although the PFC mechanism is different from the credit-based flow control mechanism, the abstract architecture in figure 1 can be reused in the implementation of PFC for RoCEv2. We observe that dual-ported Mellanox CAs can be configured to simultaneously run InfiniBand on one port and RoCEv2 on the other, implying that they must share a lot of common hardware/firmware. The *ibuf* and *obuf* modules are especially good building blocks for Ethernet NE simulation. We also continue to use the credit as a unit of available buffer space in the *ibuf* module. Because a PFC packet in [9] doesn't include available credit information, it doesn't have to be sent as frequently as in the credit-based flow control mechanism. Instead, it is sent when the buffer threshold on a port is exceeded in order to pause the other end of the VL early enough to prevent buffer overflow. More importantly, the PFC packet sending side must have enough buffer space available to store packets that might be in flight while the PFC packet is in transmission. There are two options to set a "high watermark" to trigger the PFC packet in our model, and users can select one of them. In the first option, our model uses the equation in [13] to calculate the watermark automatically. In the second option, users explicitly configure the value of this watermark.

According to [9], the PFC packet contains for each VL a pause duration whose value is based on a local estimate of when the buffer occupancy for that VL would be reduced. Further PFC packets may be necessary to refresh this duration if the situation persists. Since it is difficult to configure such an estimate, we implement the PFC pause duration as follows [13]. A large duration is specified in the PFC packet triggered by the "high watermark", and a "pause 0" packet is sent to resume traffic when triggered by a "low watermark" on that VL that is also configurable by users.

## IV. RoCEv2 Congestion Management (RCM) Implementation

Traffic flow in the current MLNX model is illustrated in Figure 2. The *gen* module is responsible for generating packets by segmenting a message received from the upper level *app* into packets with a maximum sized payload that is configurable by the user. These generated packets are passed to the *vlarb* module for VL arbitration, after which they are sent out on the wire through the *obuf* module.

If a CA's *ibuf* module receives a packet, it forwards the packet to the *sink* module which consumes the packet. If a switch's *ibuf* module receives a packet, it forwards the packet to its requested output port's *vlarb* module, which arbitrates the packet and sends it out via the *obuf* module.

To add RCM into the MLNX model, the current module architecture and connections do not need to be modified, but several key components were added.

- **1. Congestion Detection in NEs**

In RCM, NEs such as switches mark packets upon congestion. But the RCM spec [2] doesn't define the exact conditions when congestion should be detected. We implemented two kinds of congestion determination, and users can configure one for their simulation.

RCM-1a) We use the InfiniBand specification [1] definitions of the root of congestion and the victim of congestion, and the credit unit of 64 bytes for RoCEv2. Users can configure marking packets at the root alone or at both the root and the victim.

RCM-1b) We use a conventional definition of congestion, which defines the congestion state as the situation when the capacity of a requested output port is less than the sum of the traffic of the input ports competing for that output port, and packets built up in the input buffers exceed a fixed threshold value (configurable in our model). Packets will be marked at any congested points.

We monitor congestion in a NE port's *vlarb* module for each successful arbitration. If congestion is detected, the corresponding *obuf* module is set to the congestion state, which then causes it to begin to mark exiting packets.

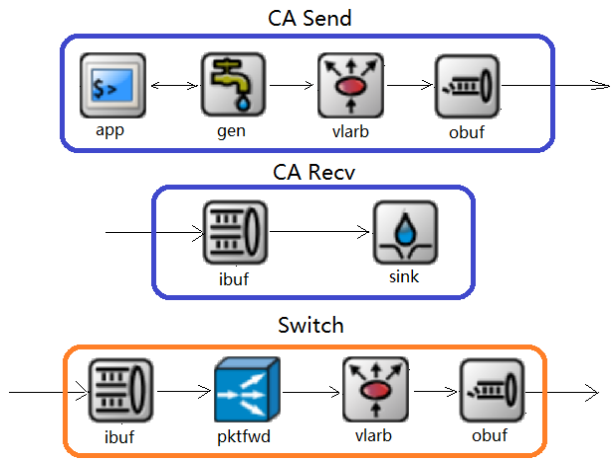

Figure. 2: Traffic Flow in the MLNX model

- **2. Packet Marking in NEs**

ECN marking simulation is done by setting a designated field in the current data packet structure (corresponding to the ECN field of the IP header).

- **3. Congestion Signaling in CAs**

When a CA's *sink* module receives a data packet with the *ECN* field set, it notifies its *gen* module which then generates a RoCEv2 CNP and sends it back to the source of the received packet via the *vlarb* module.

- **4. Injection Rate Reduction in CAs**

Upon reception of a CNP, the *sink* module notifies the *gen* module in its CA to reduce the injection rate of subsequent packets for the current connection on the specific VL. The amount of rate reduction is determined by parameters [2], but there is no guidance in the spec about how to set them. How the rate should be reduced can be configured by users. In our configuration for the example in section V we use a linear reduction strategy to implement the rate reduction. We configure $T$ to be the time to transfer a data packet with length Maximum

Transport Unit (MTU) on a wire. After receiving a CNP, the subsequent data packet won't be scheduled until $2 \times T$ has passed; after receiving two CNPs, the subsequent data packet won't be scheduled until $3 \times T$ has passed, and so forth.

- **5. Injection Rate Recovery in CAs**

A CA should increase its injection rate for a specific connection on a VL until a configurable amount of time has elapsed and/or a configurable number of bytes have been transmitted [2]. How the rate should be recovered can be configured by users. In our configuration for the example in section V each time a CA is able to recover the injection rate, it increases the rate back to the previous rate level. For instance, if the current injection rate is delayed by $3 \times T$, the CA is able to increase the rate to a delay of only $2 \times T$ for sending subsequent packets.

Both the time and the number of bytes in the recovery mechanism are also configurable in our simulator. It may require experimentation with the configuration to arrive at appropriate values that achieve optimal performance.

- **6. Congestion Statistics**

During simulation, information such as congestion duration and location are recorded. The sources and destinations of packets that flow through a congestion point are also recorded. In addition, various congestion related information such as packet latency and interval, the number of packets that are constrained by RCM and the constraint degree, etc, is collected as selected by user configuration.

## V. Performance Evaluation

Figure 3 shows the simple topology used in our experiments. Four competing traffic flows from nodes A, B, C, and D send data simultaneously to node R. Packets are transferred on the same VL, and are placed onto the links as quickly as the links can accept them, subject to the PFC. All physical links are 40 Gbps, and the packet size is configured to 2048 bytes of data. Without congestion control, these competing traffic flows suffer from the Parking Lot unfairness problem [14].

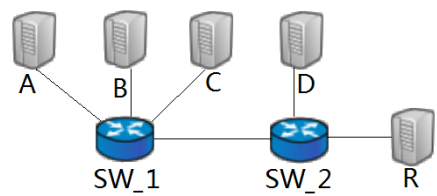

Figure. 3: Network with 4 sources simultaneously sending to dest R

We compare our simulation results with performance measurements obtained by running RoCE with PFC enabled on a hardware testbed having the topology shown in Figure 3. Switch 1 is a Mellanox SX1012, switch 2 is an Arista DCS-7050QX-32-R. Platforms A and B are equipped with Emulex Skyhawk CAs, one with firmware version 10.6.88.0, the other 10.6.144.10. Platform C is equipped with a MLNX MT27520 RoCE CA with firmware version 2.33.5100. Platforms D and R have dual-ported MLNX MT27500 CAs with firmware version 2.32.5100 that supports both InfiniBand and RoCEv2. All platforms are running OFED 3.18 [15] on Scientific Linux 7.0 with kernel version 3.10.0. Figure 4 shows the throughput comparison for each sending node in both simulation and hardware test runs. Both performance curves are almost identical, within a difference of at most 3%, verifying the accuracy of our model.

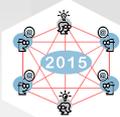

TABLE II: Simulated Throughput in Gbps in 3 RCM Scenarios

|   | NO RCM | RCM_1a | RCM_1b |
|---|--------|--------|--------|
| A | 6.3 | 9.37 | 9.29 |
| B | 6.3 | 9.42 | 9.35 |
| C | 6.3 | 9.51 | 9.43 |
| D | 18.9 | 9.72 | 9.69 |

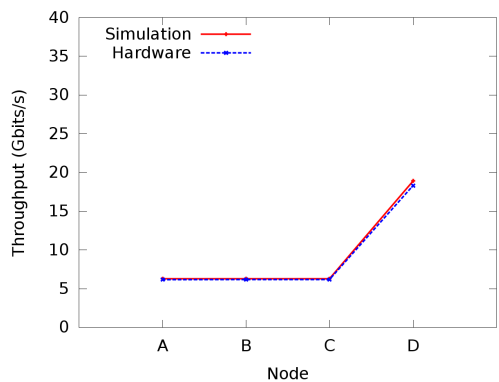

Figure. 4: Comparing simulation with RoCE hardware when PFC is enabled

Our simulations and hardware tests measure the user payload throughput observed at each node, which for a 40 Gbps link is a maximum of about 38 Gbps after accounting for protocol overhead. With PFC but without RCM[1], all input queues on both switches will fill quickly, at which point PFC will effectively reduce the injection rates of the sending nodes. Switch 2 will arbitrate equally its two input flows destined to node R, so that node D and the link from switch 1 will each see about 19 Gbps, half the maximum payload throughput of 38 Gbps seen at node R. Similarly, switch 1 will arbitrate equally its reduced output capacity of about 19 Gbps on its link to switch 2, so that flows from nodes A, B, and C are each reduced to about 6 Gbps, only one-sixth their maximum rated throughput.

Table II compares the simulated throughput of each flow in Figure 3 achieved in three scenarios: without RCM (NO RCM), RCM with the congestion detection mechanism in section IV-1a (RCM_1a), and RCM with the congestion detection mechanism in section IV-1b (RCM_1b). Without RCM, the throughput of D is three times that of each of A, B, and C due to Parking Lot unfairness, as already observed in Figure 4. With RCM_1a or RCM_1b, the four competing traffic flows share the link to receiver R almost equally. The results from RCM_1a and RCM_1b differ only slightly, indicating that these congestion detection mechanisms produce a similar effect with this simple network topology.

## VI. Conclusion

In this work, we utilize OMNeT++ to simulate the IEEE 802.1Qbb Priority-based Flow Control (PFC), and Ro-

CEv2 Congestion Management (RCM). Based on preliminary simulations, our implementation demonstrates that RCM reduces the negative effect of congestion in a simple network suffering the Parking Lot unfairness problem. Additionally, as all RoCEv2 parameters are configurable in our model, it can be easily customized and applied to any RoCEv2 cluster simulation.

## Acknowledgment

The authors would like to thank our funding agencies. This research is supported in part by National Science Foundation grants OCI-1127228, OCI-1127340, and OCI-1127341.

---

[1]RCM is not available on all our hardware CAs